# Leveraging Hyperscanning EEG and VR Omnidirectional Treadmill to Explore Inter-Brain Synchrony in Collaborative Spatial Navigation

Chun-Hsiang Chuang, Po-Hsun Peng, and Yi-Chieh Chen

***Abstract*— Navigating through a physical environment to reach a desired location involves a complex interplay of cognitive, sensory, and motor functions. When navigating with others, experiencing a degree of behavioral and cognitive synchronization is both natural and ubiquitous. This synchronization facilitates a harmonious effort toward achieving a common goal, reflecting how individuals instinctively align their actions and thoughts in collaborative settings. Collaborative spatial tasks, which are crucial in daily and professional settings, require coordinated navigation and problem-solving skills. This study explores the neural mechanisms underlying such tasks by using hyperscanning electroencephalography (EEG) technology to examine brain dynamics in dyadic route planning within a virtual reality setting. By analyzing intra- and inter-brain couplings across delta, theta, alpha, beta, and gamma EEG bands using both functional and effective connectivity measures, we identified significant neural synchronization patterns associated with collaborative task performance in both leaders and followers. Functional intra-brain connectivity analyses revealed distinct neural engagement across EEG frequency bands, with increased delta couplings observed in both leaders and followers. Theta connectivity was particularly enhanced in followers, whereas the alpha band exhibited divergent patterns that indicate role-specific neural strategies. Inter-brain analysis revealed increased delta causality between interacting members but decreased theta and gamma couplings from followers to leaders. Additionally, inter-brain analysis indicated decreased couplings in faster-performing dyads, especially in theta bands. These insights enhance our understanding of the neural mechanisms driving collaborative spatial navigation and demonstrate the effectiveness of hyperscanning in studying complex brain-to-brain interactions.**

***Index Terms*— Hyperscanning EEG, Spatial navigation, Intra-brain synchrony, Inter-brain synchrony, Functional connectivity**

## I. Introduction

SPATIAL navigation, essential for human functionality, involves accurately tracking changes in position and orientation, which are influenced by the surrounding environment and cognitive strategies [1-5]. This capability involves the use of various strategic choices. Some individuals anchor their navigation to prominent environmental features, such as landmarks, using them as cognitive aids [6-8]. Others rely on the formation of mental maps, which are detailed internal representations of space that facilitate orientation and movement [9, 10]. Although both approaches effectively enhance spatial understanding and help individuals reach their destinations, a critical question emerges: How do individuals synchronize and harmonize their mental maps with others during collaborative route planning or navigation? This question is central to understanding the mechanisms underlying collaborative spatial tasks and represents a compelling area for further research.

A frequent scenario in daily collaborative spatial tasks [11] involves guiding individuals who are unfamiliar with their surroundings to a specific destination. When such guidance is provided remotely—such as over the phone—it often relies on shared spatial understanding, which is mediated through descriptions of visual cues [6, 12] to convey their position and orientation to the guide. Moreover, this type of guidance tests the ability of the guide, who is assumed to be familiar with the setting, to recall the layout and plan routes effectively. Previous studies have indicated that an individual's familiarity with a given setting can shape how that setting is spatially conceptualized. In general, spatial navigation primarily operates within egocentric and allocentric reference frames [13]. The egocentric frame is viewer-dependent, encoding spatial relationships relative to one's own body, whereas the allocentric frame captures spatial relationships among environmental features, irrespective of the viewer's location. Existing studies [14, 15] have argued that familiar environments typically activate allocentric representations, whereas unfamiliar ones prompt an egocentric approach. Furthermore, increased familiarity with a specific location may facilitate the extraction of spatial information from both egocentric and allocentric perspectives [16]. Thus, in collaborative spatial tasks, the information provider and

This work was supported by the Ministry of Science and Technology, Taiwan (MOST 111-2636-E-007-020, 110-2636-E-007-018); National Science and Technology Council, Taiwan (NSTC 113-2636-E-007-005, 112-2636-E-007-009); and Research Center for Education and Mind Sciences, National Tsing Hua University *(Corresponding author: C.-H. Chuang).*

C.-H. Chuang is with the Research Center for Education and Mind Sciences and the Institute of Information Systems and Applications, National Tsing Hua University, Hsinchu, 30014 Taiwan (e-mail: ch.chuang@ieee.org). P.-H. Peng is with the Department of Computer Science and Engineering, National Taiwan Ocean University, Keelung, 202301 Taiwan. Y.-C. Chen is with the Research Center for Education and Mind Sciences, National Tsing Hua University, Hsinchu, 30014 Taiwan.

receiver are speculated to use different spatial reference frames during their communication. Additionally, both interacting members are likely to demonstrate flexibility, seamlessly switching between these frames to navigate effectively.

Spatial navigation involves several brain regions, including the retrosplenial cortex, posterior parietal cortex, prefrontal region, and hippocampus [17-20]. Using electroencephalography (EEG), a widely used neural imaging tool in spatial navigation research, studies have shown increased activity in the delta and theta bands within the parietal and occipital lobes during the development of mental maps and spatial navigation [21-24]. Moreover, recent neuroscientific research [25-28] has highlighted how the brain uses egocentric and allocentric reference frames for spatial navigation. For individuals who predominantly use an egocentric frame, enhanced neural activity is primarily found in the visual cortex and parieto-occipital network. Conversely, those using an allocentric perspective exhibit notable EEG activity in the occipitotemporal, bilateral inferior parietal, and retrosplenial cortical areas, which help transition between spatial reference frames [17, 29]. Based on these findings, we hypothesize that these brain regions are pivotal in collaborative spatial navigation, evident not only in individual brain activities but also in inter-brain interactions during such tasks. We also speculate the involvement of additional brain regions pivotal for social interaction tasks, such as the exchange of spatial information and coordination of mental maps, in contributing to neural synchrony. Specifically, we expect to observe increased neural synchronization in areas such as the right temporoparietal junction during tasks that require social navigation, indicating its significance in collaborative processes [30].

This study used hyperscanning technology [31, 32], a technique used to explore the brain dynamics of multiple individuals simultaneously while they engage in social interactions, to investigate how dyads cooperate in spatial navigation tasks. In particular, the interbrain synchrony (IBS)[33-35] between interacting partners, with one assigned as a leader and the other as a follower, was examined during collaborative route planning. The spatial task was conducted in a virtual reality (VR) environment with an omnidirectional treadmill to simulate real-world navigation. To analyze neural synchronization and coordination during these social interactions and their relationship with dyadic task performance, we used various intra- and inter-brain connectivity measures, namely phase locking value (PLV)[36], corrected imaginary PLV (ciPLV)[37], weighted phase lag index (wPLI)[38], and directed transfer function (dDTF)[39]. Each measure provided a unique perspective on the complex interplay of neural activities between individuals in collaborative settings. This study addressed three key research questions: 1) Do the navigation leader and follower demonstrate distinct role-specific intra-brain connectivity patterns for processes such as spatial memory and trajectory planning? 2) Do interacting dyads exhibit the inter-brain couplings crucial for managing shared attention and conflicting information during collaborative route planning? 3) Is inter-brain connectivity associated with effective task performance? These questions explore the neural dynamics underlying collaborative navigation tasks.

## II. Methods & Materials

### A. Ethics Statement

This study was approved by the Research Ethics Review Committee at National Tsing Hua University in Taiwan (number: 11001HT006). Informed consent was obtained from all participants before their inclusion in the study.

### B. Participants

A group of 70 volunteers, including 26 men and 44 women with an average age of 21.9 ± 2.8 years, participated in this study. These participants were in good health, had normal or corrected-to-normal vision, and exhibited normal hearing. The female participants were not pregnant. None of the participants had a history of chronic smoking, drug or alcohol abuse, physical impairment, psychiatric disorders, or cardiovascular issues. The participants were grouped into pairs randomly. They alternated roles as leader and follower to collaboratively complete navigation tasks.

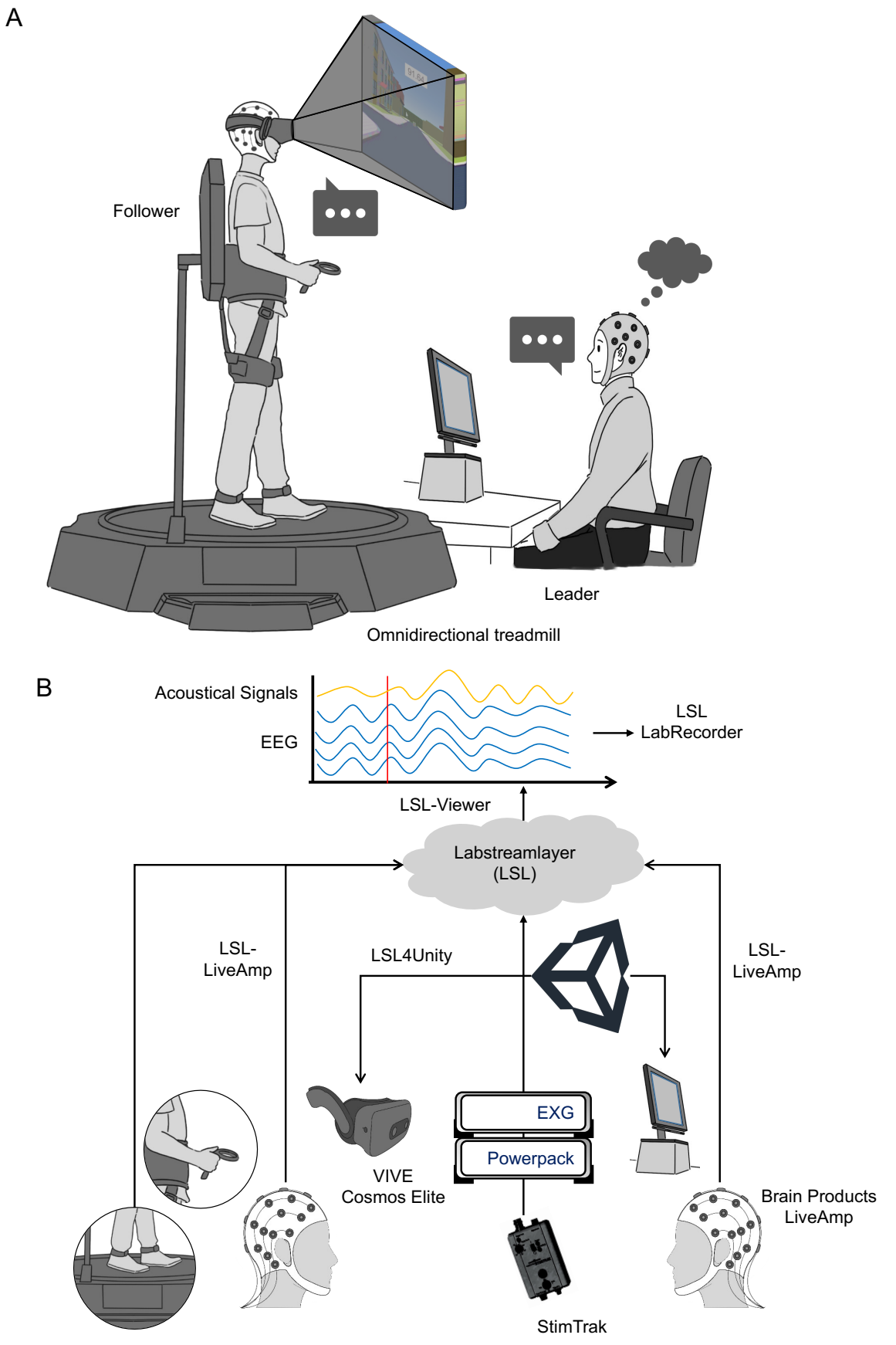

Fig. 1. Schematic representation of the experimental setup and hyperscanning EEG data collection for the dyadic route planning task. (A) The experimental configuration features one participant (follower) equipped with a VR headset and navigating on an omnidirectional treadmill, whereas a second participant (leader) monitors and directs the follower toward the target location. (B) The framework of data streaming, synchronization, and storage processes involved in the current experiment.

### C. VR Experimental Environment

In this study on dyadic route planning, participants in the follower role were deeply immersed in a VR environment that facilitated intense engagement with experimental tasks. The VR setting displayed three-dimensional (3D) street models (Fig. 1A) and was developed using the Unity engine (Unity Software Inc.). Participants navigated this virtual streetscape using a VIVE Cosmos Elite headset, enhanced by a VIVE Wireless Adapter (HTC Co., Ltd.). The inclusion of an omnidirectional treadmill (KAT Walk mini, Hangzhou Virtual & Reality Technology Co., Ltd.) and a VIVE controller not only increased participants' freedom of movement but also allowed for a realistic simulation of walking within the virtual world. The participant assigned the leader role was positioned in front of a screen throughout the experiment.

### D. VR Experimental Environment

This experiment involved a dyadic navigation task with multiple trials. One participant assumed the leader's role, guiding their partner, who undertook the follower role, to a predetermined destination. The follower navigated a VR town configured by a unique street map known only to the leader. Starting from an arbitrary point, the follower relied on the leader's cues to reach the specified endpoint. The experimental setup included maps labeled Map A and B (Fig. 3). Although both maps featured identical buildings, their layouts were different. Each participant was asked to become familiar with one map. The maps were used sequentially in the experiment. For example, the participant familiar with Map A assumed the leadership role, guiding their partner. After 30 trials, the alternate map was introduced, and the roles were reversed.

VR sickness posed challenges to the experiment's completeness and raised safety concerns. To mitigate these issues, we administered a VR sickness assessment to participants 1 month before the formal experiment. During this assessment, participants wore a VR headset and entered a 5 × 5 virtual town that closely resembled the experimental setting. Their task was to locate a specific target while exploring the virtual space for 15 minutes. We meticulously observed their physical and psychological states throughout this period. Only participants who experienced no discomfort during this preliminary test were considered suitable for the main experiment.

A week before the experiment, participants were randomly assigned one of the designated maps and instructed to familiarize themselves with it. To aid this memorization, we used a video game developed using the Unity engine, which included three training tasks. The first task was a jigsaw puzzle game where participants placed landmarks in their correct locations on a map provided for reference. The second task was a coin-searching game within the VR town, where participants navigated using cues displayed in the upper right corner of their view. The final task was similar to the first task but added a twist. In this task, participants placed landmarks based solely on their recall without the aid of the guiding map. Their gameplay was logged, and only those who engaged in this training activity at least three times and achieved perfect accuracy were selected for the main experiment. On the day of the experiment, before EEG preparation, the participants completed another round of the third task to ensure that they could flawlessly recall their maps. This task prepared them to guide their partner in the subsequent collaborative navigation activities.

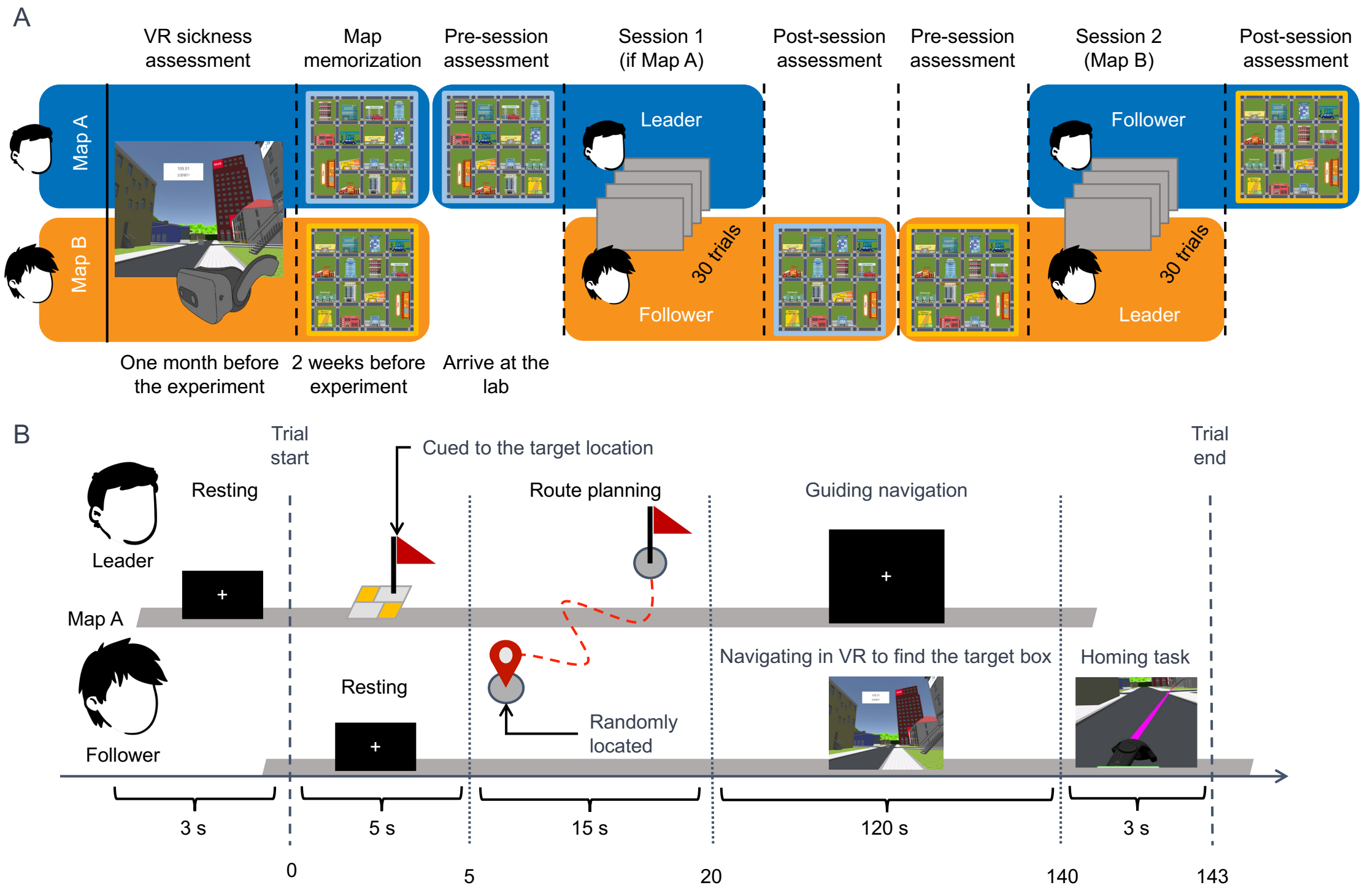

Fig. 2. Overview of the experimental paradigm for the dyadic route planning task. (A) Timeline of participant preparation phases, including VR sickness assessment and map memorization, conducted 1 month and 2 weeks before the experiment, respectively. The main experiment consists of two sessions, each featuring 30 trials of route planning and navigation, with pre- and post-session evaluations for both the leader and the follower. (B) A detailed step-by-step breakdown of a single trial illustrating the progression of activities for the dyads, from resting state to cue presentation, route planning, and navigation guidance, and culminating in the homing task, with durations specified for each phase.

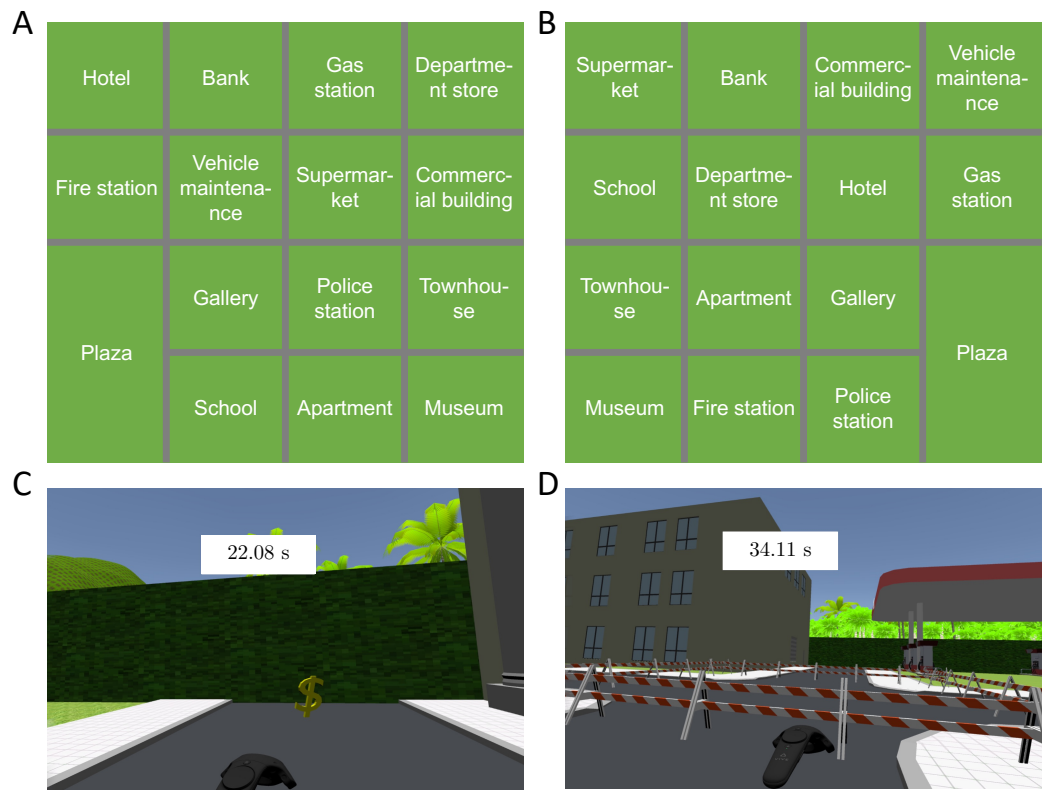

Fig. 3. Map layout and virtual environment for the dyadic navigation task. (A) Two grid layouts display the distribution of various buildings and landmarks used in this study. (B) Screenshots from the VR scenes, with one showing navigation toward a target and the other depicting an obstacle encountered along the path.

As illustrated in Fig. 3A and 3B, the two street maps designed for this study measured 4 × 4 and featured 15 building models, including a police office, commercial building, school, gallery, and museum. One of the two maps was randomly selected for the first session, with the other used for the second session. The participant familiar with the chosen map assumed the role of the leader and sat next to the follower who wore a VR headset and stood on an omnidirectional treadmill. Each session consisted of 30 trials and involved locating a coin-containing box in a virtual town (Fig. 3C).

At the beginning of each trial, a 5-second block was initiated. During this period, the follower saw a fixed white cross against a black background, whereas the leader viewed an image indicating the target box's location. Upon the disappearance of the fixation and cue image, the follower was randomly placed at an unspecified location in the virtual town. This was followed by a 15-second discussion block dedicated to route planning. During this phase, the follower described the nearby buildings to reveal their location. The leader then informed the follower of the target box's position. Together, they strategized the best path from the starting to ending points.

After planning, the follower began navigating the virtual town in search of the target box. Given the unfamiliarity of the streets, the pair continued their conversation to ensure that the follower was effectively guided toward the target. Roadblocks (Fig. 3D) were introduced in half of the trials to add complexity to the task, necessitating detours. In addition, decoy boxes were scattered throughout the town as potential distractions.

If the task duration exceeded 120 seconds, the current trial would be terminated, and the subsequent one would commence. Upon reaching the destination, the follower underwent a homing direction test [40] to identify the initial starting point using a VR controller that emitted a laser beam within the virtual space. Both the follower's navigation route and homing were captured using LSL2Unity [41]. After the completion of all experimental trials, the follower participated in a jigsaw puzzle game to assess their familiarity with the unfamiliar map layout. The achieved score was then documented.

## E. Hyperscanning EEG

In this hyperscanning EEG study, we simultaneously captured EEG signals from each dyad using two identical LiveAmp wireless amplifier systems, each equipped with 32-channel actiCAP slim active electrodes from Brain Products GmbH, Germany. These electrodes were mounted on the EEG cap and adhered to the extended 10-20 system layout (Fp1, Fz, F3, F7, FT9, FC5, FC1, C3, T7, TP9, CP5, CP1, PZ, P3, P7, O1, OZ, O2, P4, P8, TP10, CP6, CP2, CZ, C4, T8, FT10, FC6, FC2, F4, F8, and FP2). A ground electrode was placed at Fz, and the reference electrode was situated at FCz. To ensure optimal signal acquisition, conductive gel (SuperVisc, EASYCAP Gmbh) was used to fill the space between the electrodes and the scalp. The impedance for every electrode was ensured to be under 10 kΩ. EEG data were captured at a sampling rate of 1,000 Hz, with an ADC resolution of 16 bits.

Consistent with the configurations in our previous study [42],

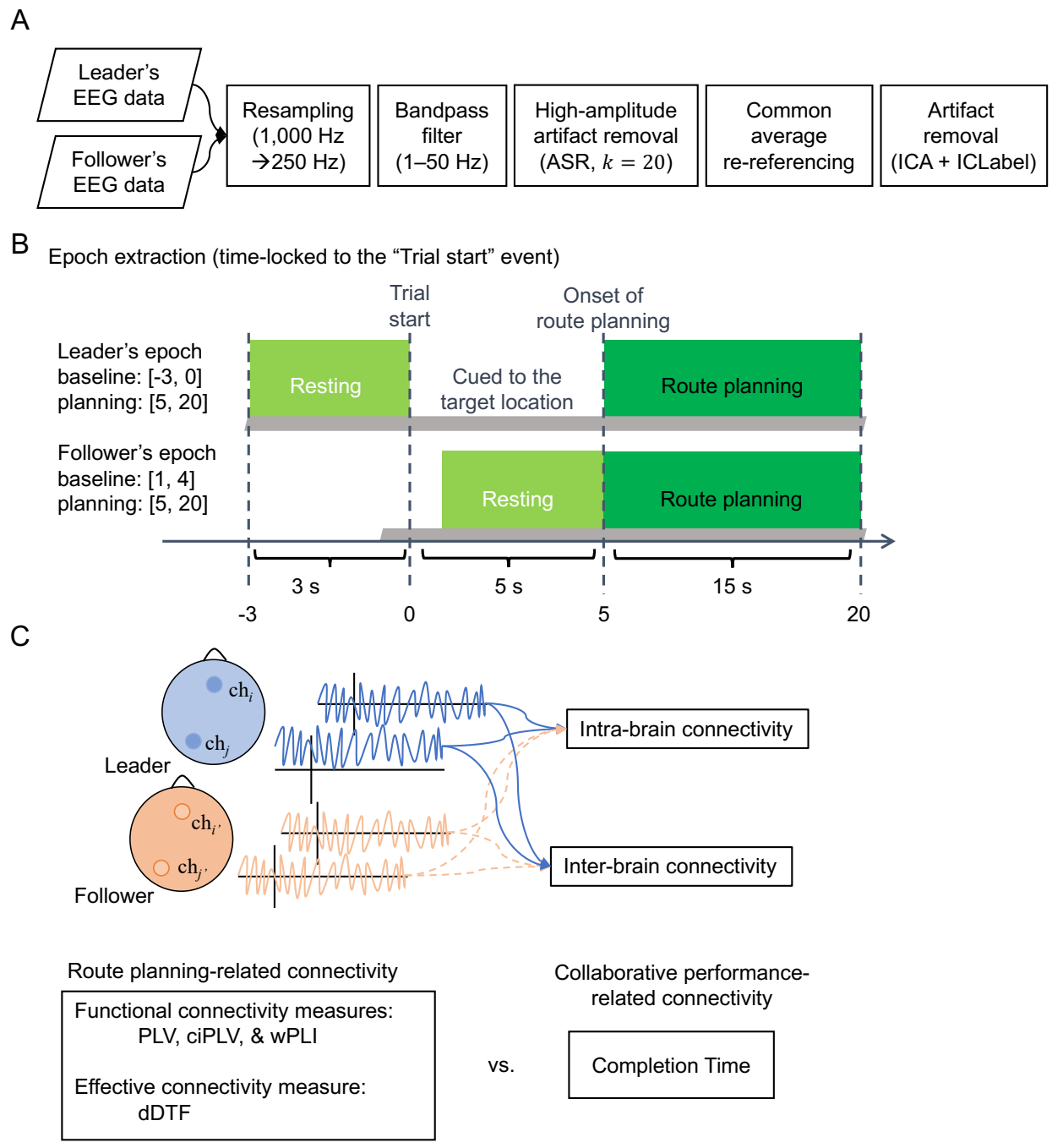

Fig. 4. EEG data processing pipeline and analysis for a study on collaborative route planning. (A) Initial processing steps for EEG data from both the leader and follower. (B) Epochs are extracted time-locked to the "Trial start" event, with specific pre-trial baselines and route planning intervals defined for the leader and follower. (C) Analyses of intra-brain and inter-brain connectivity are integrated with evaluations of collaborative performance, which include behavioral data and task completion times.

we used Labstreaminglayer [41], integrating LSL-LiveAmp, LSL4Unity, BrainVision LSL Viewer, and LSL-LabRecorder, to construct a data acquisition platform (Fig. 1B). This platform was adept at seamlessly synchronizing data streams and event markers from both Unity and dual EEG devices. Real-time monitoring of all data streams was conducted, safeguarding against potential equipment malfunction. The collected data were saved in an Extensible Data Format file. Subsequently, we used the MoBILAB EEGLAB plugin to transform the data into a format compatible with EEGLAB [43].

### F. Data Processing and Brain Connectivity Measures

From the behavioral data stream and associated event markers, we recorded the walking trajectory, completion time (i.e., latency for each target search trial), and homing angle. To compare the complexity of the two maps, we used two metrics: the number of turns and the taxicab distance required for an optimal route in each trial.

EEG signals were mainly processed using native and plugin functions from EEGLAB version v.14.1.2 [43] and Matlab R2020a, adhering to Makoto's preprocessing guidelines [44]. As illustrated in Figure 4A, raw EEG data were initially downsampled to 250 Hz and then band-pass filtered between 1 and 50 Hz. Line noises were mitigated using the cleanline EEGLAB plugin [45]. The data were further refined through artifact subspace reconstruction with a parameter ($k$) value of 20 [46] to eliminate prominent artifacts, such as eye movements, blinks, and other motions. Finally, the data were processed using common average referencing.

This study examined the synchronization of mental maps within leader–follower dyads during collaborative route planning. Specifically, we evaluated brain-to-brain coupling during the 15-second discussion block allocated for route strategy. In this phase, the follower identified nearby landmarks to communicate their location, whereas the leader formulated the most effective path to reach the target location. As shown in Fig. 4B, EEG epochs within the 15-second route planning phase were extracted for subsequent analysis. Corresponding baseline segments were drawn from the initial resting-state blocks for comparison.

Three measures of functional connectivity, namely PLV[36], ciPLV [37], and wPLI [38], were used to examine intra- and inter-brain connections. In addition, dDTF [39] was used to assess causality within and between brains, providing a comprehensive view of functional and effective connectivity (Fig. 4C). These connectivity analyses were conducted across a spectrum of EEG frequency bands, including delta (1–3 Hz), theta (4–7 Hz), alpha (8–12 Hz), beta (13–30 Hz), and gamma (31–50 Hz). Moreover, we determined the relationship between intra-/inter-brain couplings and task performance metrics (follower's post-test scores and completion times).

TABLE I
DEMOGRAPHICS AND ASSESSMENT SCORES FOR DIFFERENT MAPS

| | Map A | Map B | Statistics | *p* |
|---|---|---|---|---|
| | n = 70 | | | |
| Women (Men) | 21 (14) | 23 (12) | $x^2(1)$=0.25 | .621 |
| Age (years) | 22.1 ± 2.1 | 21.7 ± 1.8 | $t(68) = -0.79$ | .433 |
| Dyadic sex composition | | | | |
| Female-Female, Female-Male, Male-Male | 16, 12, 7 | | | |
| Assessments | | | | |
| Map memorization task (%) | 100.0 ± 0.0 | 100.0 ± 0.0 | N/A | N/A |
| Pre-session assessment (Leader, %) | 100.0 ± 0.0 | 100.0 ± 0.0 | N/A | N/A |
| Post-session assessment (Follower, %) | 56.9 ± 27.1 | 63.7 ± 29.4 | $t(68) = -1.01$ | .314 |
| Completion rate (%) | 99.2 ± 0.2 | 98.4 ± 0.2 | $t(68) = -0.79$ | .435 |
| Completion time (second) | 52.8 ± 10.9 | 50.8 ± 10.2 | $t(68) = -0.80$ | .425 |

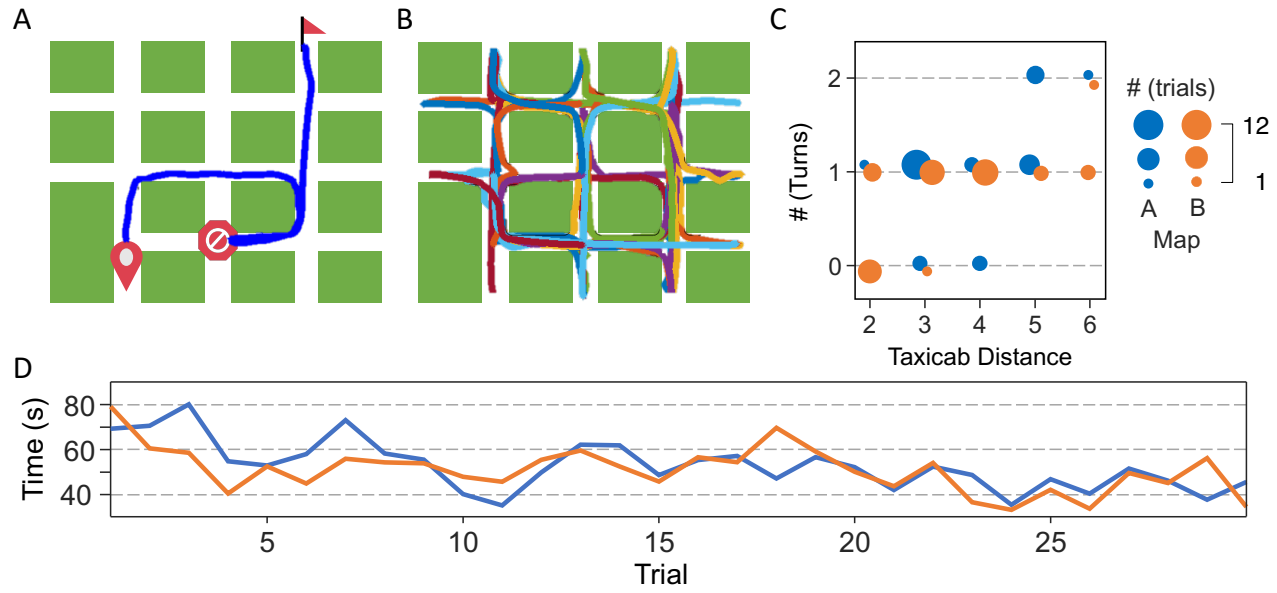

Fig. 5. Different aspects of navigational behavior and complexity in the navigation task. (A) Trajectory of a single trial, highlighting a detour taken. (B) Aggregation of trajectories across 30 trials. (C) Comparative analysis of the complexity of Maps A and B using two measures, taxicab distance and number of turns, with the size of dots representing the number of trials. (D) The average completion times over all trials for both maps, indicating the efficiency of navigation over repeated attempts.

## III. RESULTS

### A. Behavioral Results

Figure 5A illustrates an example of successful navigation around a roadblock, depicting the chosen detour path. Figure 5B confirms that the paths taken in all trials were systematically recorded, providing a comprehensive dataset of navigational choices and strategies. Participants engaged with one of the two maps used in the study, labeled Map A and Map B. As shown in Figure 5C, no significant differences were observed in the taxicab distance and the number of turns between the two maps ($p$>.05). This finding suggests that the complexity of the two maps was statistically comparable for the purpose of the investigation. As depicted in Figure 5D, as dyads became more familiar with the task, their completion times improved, indicating a learning effect. This trend of increased navigation proficiency was observed in both maps, demonstrating consistent improvement in performance across different navigational challenges.

Demographic analysis revealed an equal distribution of sex across both map groups, with 21 women and 14 men for Map

TABLE II
DEMOGRAPHICS AND ASSESSMENT SCORES OF TWO GROUPS WITH SLOW & FAST TASK COMPLETION TIMES

| | Slow | Fast | Statistics | *p* |
|---|---|---|---|---|
| Age (years) | 21.4±1.5 | 22.5±2.2 | $t(68)=2.32$ | .023 |
| Dyadic sex composition | | | | |
| Leader (F, M) | 23, 12 | 21, 14 | | |
| Follower (F, M) | 22, 13 | 22, 13 | | |
| FF, FM, MM, MF | 16, 7, 6, 6 | 16, 7, 8, 6 | | |
| Assessments | | | | |
| Map memorization task (%) | 100.0±0.0 | 100.0±0.0 | N/A | N/A |
| Pre-session assessment (Leader, %) | 100.0±0.0 | 100.0±0.0 | N/A | N/A |
| Post-session assessment (Follower, %) | 61.1±29.2 | 59.4±27.8 | $t(68)=-0.25$ | .802 |
| Task performance | | | | |
| Completion rate (%) | 98.4±2.5 | 99.7±1.2 | $t(68)=2.85$ | .006 |
| Completion time (second) | 59.7±7.6 | 41.9±5.8 | $t(68)=-11.09$ | <.001 |

A and 23 women and 12 men for Map B. As shown in Table I, statistical tests confirmed that the difference in sex distribution was not significant ($\chi^2(1)=.25$, $p=.621$). In addition, age comparisons showed no significant disparity, with an average of 22.1 (SD=2.1) years for Map A and 21.7 (SD=1.8) years for Map B ($t(68)=.79$, $p=.433$).

In terms of dyadic sex composition, the participants were grouped into 16 female–female pairs, 7 male–male pairs, and 12 mixed-sex pairs. All participants performed exceptionally well in the map memorization task and pre-session assessment, achieving a 100% success rate. This finding indicated no significant difference in map learning between the two conditions.

In terms of performance post-session, followers scored 56.9% (SD=27.1) for Map A and 63.7% (SD=29.4) for Map B. Statistical analysis showed no significant difference between these scores ($t(68)=-1.01$, $p=.314$). The completion rates for both Map A and Map B were nearly perfect and comparable, with scores of 99.2% (SD=.2) and 98.4% (SD=.2), respectively, and this similarity was statistically supported ($t(68)=-.79$, $p=.435$). The average completion times for navigating Map A and Map B were 52.8 (SD=10.9) seconds and 50.8 (SD=10.2) seconds, respectively. Statistical analysis indicated no significant difference in navigation times between the two maps ($t(68)=-.80$, $p=.425$).

### B. Task Completion Time

Table II presents demographic data and assessment scores for two groups categorized by their task completion time—slow and fast. The slow group consisted of 23 women and 12 men in the leader role and 22 women and 13 men in the follower role. The fast group included 21 women and 14 men in the leader role and 22 women and 13 men in the follower role. In terms of dyadic sex composition, each group had a mix of female–female, female–male, male–male, and male–female pairs, with similar distributions of female and male participants across both leader and follower roles. The results revealed that both the slow and fast groups performed equally well in the map memorization task and the pre-session assessment, achieving a perfect score. The post-session assessment scores for followers did not significantly differ between the groups (61.1 ± 29.2 for the slow group vs. 59.4 ± 27.8 for the fast group, $t(68)=-.25$, $p=.802$). However, task performance differed significantly, with the fast group achieving a higher completion rate (99.7%±1.2%) than the slow group (98.4%±2.5%; $t(68)=2.85$, $p=.006$). Completion times were significantly shorter for the fast group, averaging 41.9 (SD=5.8) seconds, than the slow group, averaging 59.7 (SD=7.6) seconds.

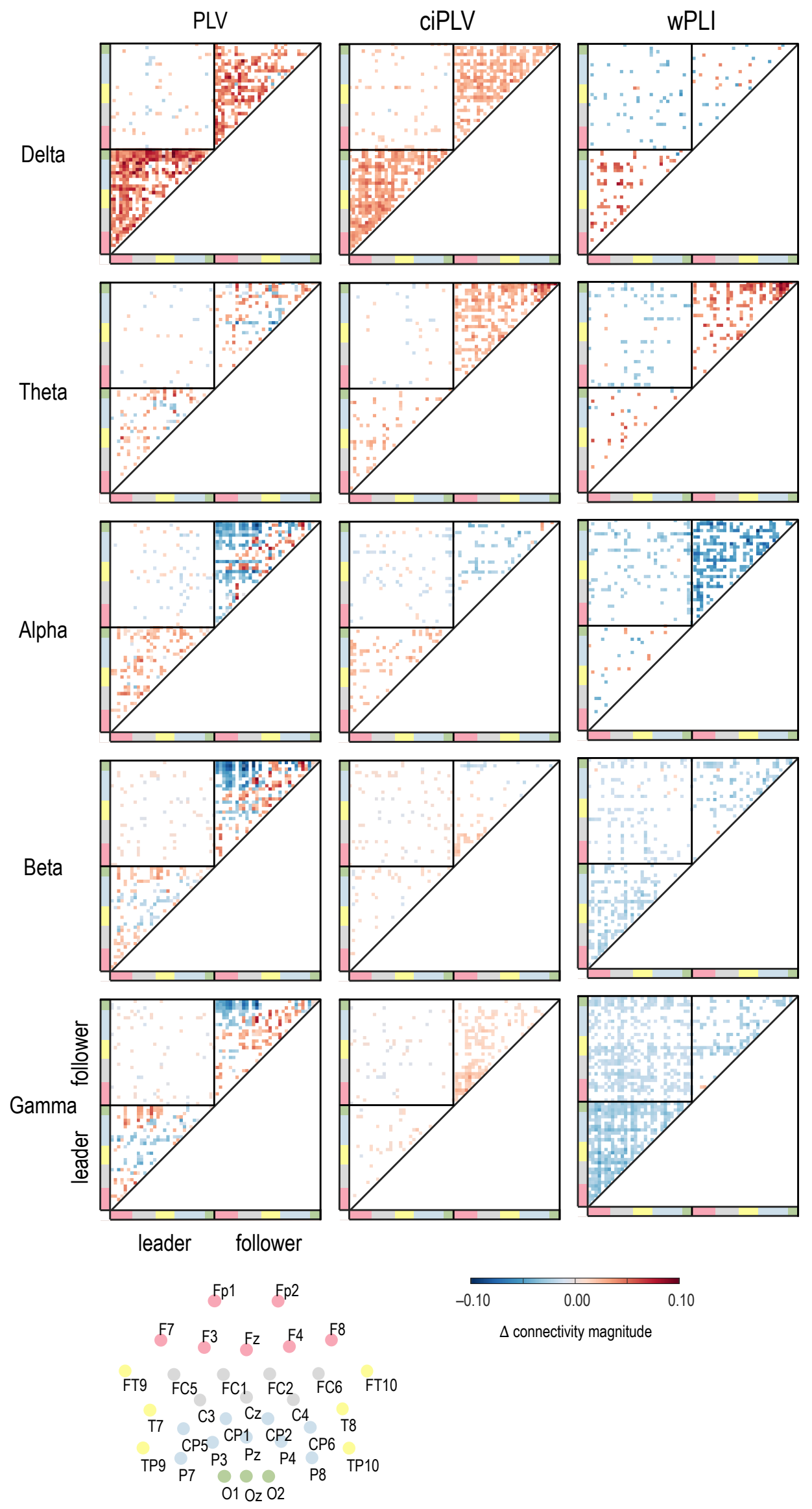

Fig. 6. Intra- and inter-brain connectivity in route planning. Each subfigure illustrates the connectivity magnitude, quantified by PLV, ciPLV, or wPLI, in a specific frequency band. In each subfigure, the right-upper and left-lower triangles represent the intra-brain connectivity of leaders and followers, respectively. The left-upper square shows inter-brain connectivity between dyadic pairs.

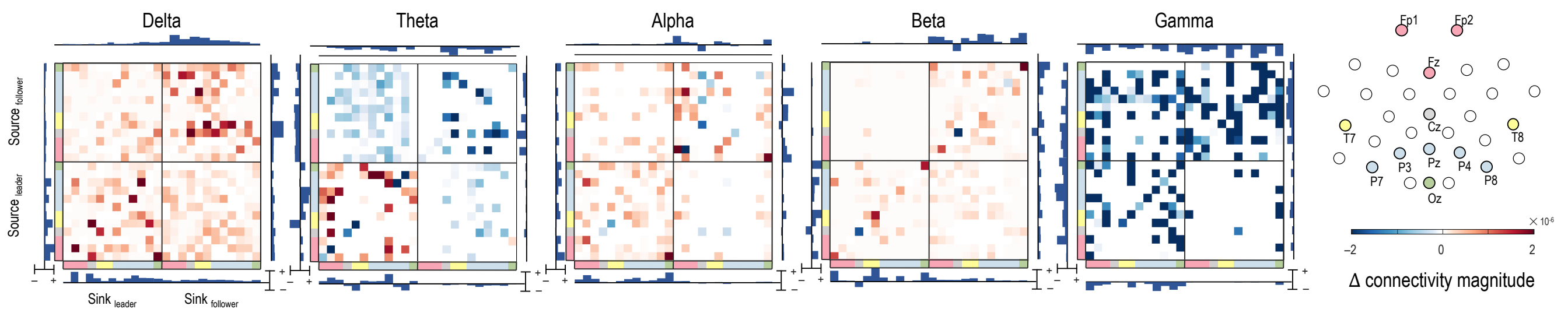

Fig. 7. Causal brain connectivity patterns within and between leaders and followers across various EEG frequency bands, as estimated by dDTF. The matrices are segmented into four quadrants: the third and first quadrants display intra-brain effective connectivity for leaders and followers, respectively, whereas the fourth and second quadrants exhibit the inter-brain connectivity from leaders to followers and from followers to leaders, respectively. The surrounding bars indicate the sum of causal magnitudes, with vertical bars representing the total outgoing (source) connectivity and horizontal bars representing the incoming (sink) connectivity. The accompanying electrode map provides a reference for scalp locations, and significant connectivity is denoted in color.

Statistical analysis indicated a significant difference in completion times between the groups ($t(68)=-11.09$, $p<.001$). This finding suggests that while both the groups had a similar initial understanding of the task, the fast group was more efficient in task execution.

## C. Functional Intra- and Inter-brain Connectivity

Figure 6 depicts the functional intra-brain connectivity of leaders and followers and inter-brain connectivity between dyadic pairs measured using PLV, ciPLV, and wPLI across five frequency bands. The figure provides a detailed comparison of brain connectivity within and between individuals engaged in route planning for collaborative navigation tasks, highlighting distinct neural patterns associated with each role and the synchrony between partners.

In the delta band, both PLV and ciPLV revealed a significant increase in intra-brain connectivity in most brain regions in both the leaders and the followers, indicating enhanced neural coordination within each individual during the task. The wPLI data showed a greater increase in intra-brain connectivity in the leaders, indicating a more significant role of leader-specific neural processes during the task, as measured by this particular connectivity index. In the theta band, ciPLV and wPLI indicated increased intra-brain connectivity in the follower, especially between the parietal and occipital regions. This finding suggests a more active engagement of these areas in followers. In the alpha band, the intra-brain connectivity of the follower was significantly reduced when compared with that of the leaders. This reduction, particularly identified by the PLV, was predominantly observed between the occipital and parietal regions. This reduction was more widespread across various brain regions when measured using the wPLI. Conversely, both PLV and ciPLV indicated an increase in intra-brain connectivity in the leaders, highlighting differing patterns of neural engagement between the two roles within this specific band.

In the follower, connectivity patterns in the beta and gamma bands, as estimated using PLV, were similar to those observed in the alpha band. The PLV highlighted increased intra-brain connectivity in the follower across the central, temporal, and parietal regions in the alpha, beta, and gamma bands. The observed results for the leaders appeared to contrast with those for the followers, indicating an inverse relationship in their connectivity patterns, particularly in beta and gamma bands across the examined frequencies. The ciPLV indicated that increased connectivity was mainly noted in the frontal regions in the follower. Meanwhile, the wPLI revealed decreased connectivity in the alpha band for the follower and in the beta and gamma bands for both roles, highlighting distinct neural engagement strategies between leaders and followers.

Although there was a clear reduction in inter-brain connectivity in the gamma band, drawing definitive conclusions regarding inter-brain connectivity using the three functional connectivity measures proved challenging.

## D. Inter- and Intra-Brain Causality Associated with Task Performance

Figure 7 illustrates the distinct patterns of intra- and inter-brain causality within and between leaders and followers across various EEG frequency bands, as quantified by the dDTF. This visual representation highlights the directional flows of neural communication that are crucial during route planning.

In the delta band, we noted a significant enhancement of intra-brain causality in most regions in both leaders and followers, depicted in red in Fig. 7). This finding indicates an association of increased connectivity with route planning. In addition, inter-brain causality was markedly enhanced from leaders to followers, spanning across multiple brain regions in both individuals. Similar modest enhancements in both intra- and inter-brain causality were observed within the alpha and beta frequency bands, with the beta band showing a particularly rare occurrence of follower-to-leader inter-brain causality.

In the theta band, we noted a marked increase in intra-brain causality in the leaders, predominantly in the parietal regions. Conversely, the follower exhibited decreased intra-brain causality. This reduction pattern was particularly evident in the gamma band in both the leader's and follower's brain activity as well as in the connectivity from the follower to the leader.

Figure 8 depicts the neural dynamics associated with task duration in the collaborative navigation task, specifically focusing on differences in theta band connectivity between the fast and slow groups. This figure shows changes in neural connections both within individual brains and between team members. The results indicate that the intra-brain connectivity of followers may be a crucial factor in task completion. Moreover, we observed that decreased connectivity in specific brain regions was associated with more efficient navigation performance. The findings on inter-brain connectivity suggest that the fast group exhibited decreased neural coupling between dyads. This reduction in neural communication, particularly from the leader's to the follower's parietal regions and vice versa, may be correlated with more efficient task completion in the fast group.

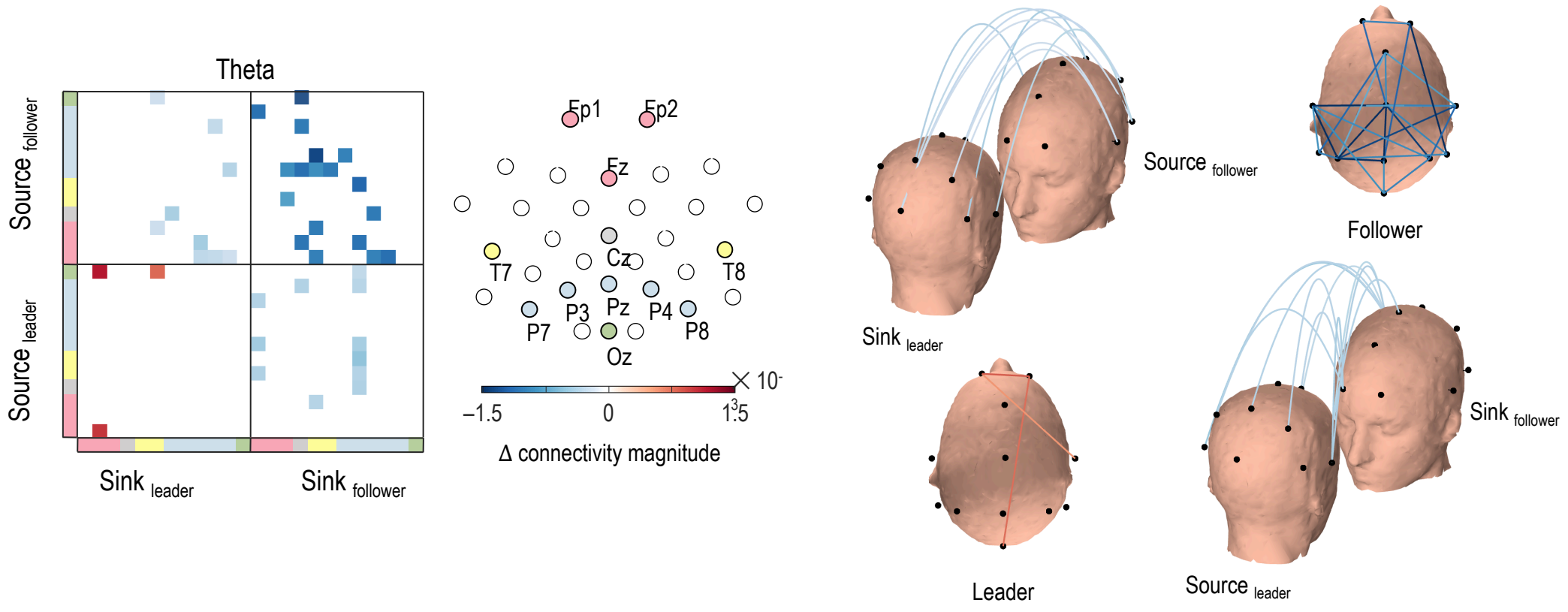

Fig. 8. Neural dynamics correlate with performance speed in collaborative tasks. This figure illustrates differences in theta band connectivity between groups with fast and slow task completion times. The 3D brain models and matrix reveal changes in intra- and inter-brain connection. Connectivity shifts are color-coded, with red indicating increased connections and blue showing decreases in the fast task completion group compared with the slow task completion group.

## IV. Discussion

Sharing spatial information involves more than just noting location changes; it requires fostering mutual understanding among individuals to efficiently exchange spatial details. Moreover, creating shared cognitive maps enhances collaboration in various contexts, particularly in guiding others through unfamiliar areas. This collaborative dynamic is fundamental for effectively navigating and working together in diverse settings. In this study, unique map layouts were designed to create asymmetrical knowledge among participants in a dyad, assigning specific roles of leader and follower. In the collaborative navigation task, interactions between leaders and followers included critical elements, such as a sense of agency, perspective-taking, shared attention, and the analysis of spatial memory processing over time, which are essential for successful cooperation. This setup enabled the exploration of how varying levels of spatial information affect collaborative interactions and their associated neural connectivity, both inter- and intra-brain, during navigation tasks.

Previous studies have observed significant alpha suppression in the temporal and parietal areas during spatial navigation tasks. This finding can be attributed to the need for increased cortical excitability essential for spatial updating [5]. Regardless of specific brain regions involved in encoding and retrieving spatial data, the alpha neural network plays a crucial role in processing spatial reference frames [47]. This study builds on these insights by showing decreased alpha couplings among brain regions, particularly in followers. This role-specific alpha decoupling likely reflects the followers' need to integrate their location with guidance from leaders, manage conflicting information, and facilitate the transformation of spatial information.

Effective communication and strategic planning are essential in route planning tasks, where participants must share spatial information efficiently. Delta oscillations, which are linked to attention-demanding encoding, semantic processing, and speech perception, play a significant role in these activities [48, 49]. The findings of this study regarding both functional and effective connectivity are consistent with those of previous studies on conversational interactions and speech content tracking, highlighting the cognitive processes involved in navigation tasks. Although previous studies [50] have shown that increased theta activity in medial–temporal–parietal regions correlates with better navigation performance, our findings revealed a decrease in both intra-brain connectivity within followers and inter-brain connectivity between leaders and followers in groups that completed tasks quickly. This result contrasts with that of a previous study [51] that measured task performance using millisecond-scale reaction times, suggesting that the need for attention resources is brief and transient in nature. The findings of this study stem from route planning sessions where dyadic participants discussed their spatial strategies. Notably, increased theta connectivity was observed in leaders, aligning with their involvement in trajectory planning [52] and place memory [53], whereas a decrease was noted in followers. This divergence in brain activity, tailored to each role's demands, led to decreased inter-brain theta couplings. These results suggest that distinct cognitive processes are at play within the dyad during collaboration, highlighting role-specific neural dynamics.

The current experimental setup clearly designated participants as leaders or followers, creating an asymmetry in spatial information that naturally established leadership roles. This structure aligns with the findings of previous research[54], which observed increased alpha inter-brain effective connectivity from followers to leaders but not vice versa, supporting the notion that leaders and followers engage differently in collaborative tasks. However, upon deeper analysis, the designated roles of leader and follower in the experiment might not accurately reflect actual leadership dynamics. The observed alpha suppression in followers suggests that they play a more critical role in driving the interaction [55]. This pattern indicates that followers could be the ones leading the navigational decision-making process, challenging the traditional perception of their role.

Previous studies [56, 57] have demonstrated that the right temporoparietal junction is crucial in social interactions. This association is indicated by strong inter-brain synchronization observed in the interacting dyad, as noted in functional near-infrared spectroscopy signals during tasks involving social navigation [30]. The findings of this study highlight that the temporal and parietal areas are crucial functional hubs in collaborative route planning. Moreover, the results revealed an inverse relationship where decreased inter-brain synchronization in these regions is correlated with more efficient task execution, suggesting that less neural coupling enhances performance. This phenomenon may be explained by irruption theory [58], which posits that higher subjective engagement and motivation in social interactions lead to increased neural entropy, resulting in lower inter-brain synchronization in the fast task completion group. Another possible explanation for the observed results is the influence of subjective task difficulty on inter-brain synchrony. A previous study [59] demonstrated decreased inter-brain synchrony among participants engaged in easier tasks compared with those involved in more challenging tasks. In this study, the fast task completion group appeared to have established efficient communication, which facilitated seamless sharing of spatial information and intentions. This likely resulted from a tacit understanding among team members. The concise and brief nature of their communication during planning phases suggests that these interacting members perceived the task as easier.

Although the brain's navigation networks can distinguish between different spatial reference frames used during navigation [40], this study did not investigate how these frames affect intra- or inter-brain connectivity. This omission was due to the variability in the use of spatial reference frames, which can significantly change based on task demands. Factors such as cultural background [60], the nature of verbal descriptions [61], the dimensionality and spatial scale of the environment [62], and specific characteristics of navigation tasks [63] all influence the choice of spatial strategies employed. This variability suggests that spatial navigation strategies are not one-size-fits-all but are instead adapted to meet the unique requirements of different contexts and challenges. Our study did not include a reference frame proclivity test because the experimental setup involving complex person-to-person interaction and the nature of the environment did not lend itself to straightforward categorization into a single spatial strategy. Participants were expected to flexibly use egocentric, allocentric, and beacon strategies, reflecting the multifaceted approach observed in real-world navigation where strategies are dynamically integrated as needed. This complexity opens avenues for future research to explore how different spatial strategies specifically influence neural connectivity during navigation tasks.

## V. Conclusions

This study investigated the neural mechanisms underlying collaborative spatial navigation by using hyperscanning EEG technology to examine intra- and inter-brain connectivity among dyads engaged in route planning. The findings revealed role-specific neural strategies, with leaders exhibiting increased connectivity in frontal and parietal regions indicative of active planning and decision-making. In contrast, followers displayed different connectivity patterns that aligned with their roles in the navigation task. In addition, our analysis of inter-brain connectivity provides new insights into synchronization processes between individuals, enhancing our understanding of collaborative behavior from a neuroscientific perspective. Moreover, we observed notable differences between fast and slow task completion groups in collaborative spatial tasks. This variance in connectivity not only indicates the importance of neural synchronization for task efficiency but also highlights that quicker task completion may be associated with more optimized neural communication strategies. These insights expand our understanding of the neural basis of collaborative performance, emphasizing the critical role of theta band activity in facilitating efficient teamwork and decision-making processes in spatial navigation tasks.